\documentclass[showpacs,aps,prd,nofootinbib,floatfix,amsmath,amssymb]{revtex4}
\usepackage{graphicx}
\usepackage{multirow}
\newcommand{\uu}{{\mathcal{U}}}

\begin{document}

\title{Weak properties of the $\tau$ lepton via  a spin-0 unparticle}

\author{ A. Moyotl}
\email[E-mail:]{amoyotl@sirio.ifuap.buap.mx}
\affiliation{Instituto de F\'isica,  Benem\'erita Universidad
Aut\'onoma de Puebla, Apartado Postal 72570 Puebla, M\'exico}
\author{G. Tavares-Velasco}		
\affiliation{Facultad de
Ciencias F\'\i sico Matem\'aticas, Benem\'erita Universidad
Aut\'onoma de Puebla, Apartado Postal 1152, Puebla, Pue., M\'
exico}

\date{\today}

\begin{abstract}
We consider a spin-0 unparticle and calculate its contribution to the weak properties of a fermion, which in the proper limit reduce to previously reported results for the electromagnetic properties. We then obtain an estimate for the electromagnetic and weak properties of the $\tau$ lepton. For the unparticle parameters, we consider the most recent bounds from  the lepton flavor violating decay $\tau\to 3\mu$, the muon anomalous magnetic moment, and the data for monojet production plus missing transverse energy at the LHC. In the most promising promising scenario, it is found that the unparticle contributions to the $\tau$ electromagnetic properties can be larger than the contributions predicted by the standard model (SM)  and some of its extensions, such as the SeeSaw model and  extensions of the minimal supersymmetric standard model (MSSM) with a mirror fourth generation and vectorlike multiplets. As for the contributions to the weak properties, they are larger than the SM contributions but are much smaller than the current experimental limits. We also discuss the case of the electromagnetic and weak properties of the bottom and top quarks.
\end{abstract}

\pacs{13.40.Em, 12.60.-i}

\date{\today}

\maketitle
\section{introducci\'on}

The theoretical and experimental study of the electromagnetic properties of elementary particles have long represented an interesting research topic in particle physics. Along these lines, the study of the magnetic dipole moment (MDM) and the electric dipole moment (EDM) of fermions has attracted considerable attention in the literature. In particular, it is believed that the study of the EDM may shed light on new sources of CP violation. On the other hand, less attention has been paid to the weak properties of fermions, the weak magnetic dipole moment (WMDM) and the weak electric dipole moment (WEDM),  which are the analogues of the fermion electromagnetic properties but are associated with the interaction of a fermion with the neutral weak gauge boson.  In  the standard model (SM), the EDM and the WEDM arise from the CP-violating phase appearing in the  Cabibbo-Kobayashi-Maskawa (CKM) matrix \cite{Kobayashi:1973fv}. Although such a phase is enough to explain the CP  violation observed in the $K^0-\bar{K}^0$ system \cite{Christenson:1964fg},  it does not account for the baryogenesis of the universe.  However, recent evidences of neutrino oscillations \cite{osci} suggest that these particles have nonzero mass, which opens up the possibility for lepton flavor violation (LFV) and a source of CP violation in the lepton sector.

Although the electron and muon electromagnetic properties
have been measured with high accuracy, our knowledge of the $\tau$ electromagnetic properties is still beyond an acceptable level, which is due mainly to the fact that the $\tau$ lifetime is too short to allow one to directly measure its interaction with an electromagnetic field. However, indirect bounds on the $\tau$ electromagnetic properties have been obtained via the study of the deviations of the cross sections for $\tau$ production at the CERN LEP. For instance, the current constraints on the $\tau$ electromagnetic properties were obtained through  the study of the processes  $e^+e^- \to \tau^+\tau^- \gamma$ and $e^+e^- \to e^+e^-\tau^+\tau^-$. Limits on the latter reaction allowed the DELPHI collaboration \cite{Abdallah:2003xd} to place the following bounds:

 \begin{eqnarray}
-0{.}052< &a_\tau& < 0{.}013, \\
-0{.}22< & \text{Re}(d_\tau) &< 0{.}45,\\
-0{.}25 <& \text{Im}(d_\tau) &< 0{.}008,\label{etau}
\end{eqnarray}
where the EDM is expressed in units of $10^{-16}$ e cm. These results are well beyond the theoretical predictions of the SM: $a_\tau^{\text{SM}}=1177{.}21(5)\times10^{-6}$\cite{asm} and $d_\tau^{\text{SM}}<10^{-34}$ e cm \cite{dsm}. On the other hand, the weak properties of the $\tau$ lepton remain almost unexplored up to date, though the first constraints on them were obtained from the study of the cross section for the process $e^+e^- \to \tau^+\tau^-$ by using a center-of-mass energy near the $Z$ resonance \cite{Acciarri:1998zc,Acton:1992ff}. The current bounds on the weak properties of the $\tau$ lepton, which were obtained by the  ALEPH collaboration \cite{Heister:2002ik} using a data sample collected from 1990 to 1995 corresponding to integrated luminosity of 155 pb$^{-1}$, are given by:

\begin{eqnarray}
\text{Re}(a_\tau^{W}) &<& 1{.}1\times10^{-3},\\
\text{Im}(a_\tau^{W})&< & 2{.}7 \times 10^{-3},\\
\text{Re}(d_\tau^{W}) &<& 0{.}5 \times 10^{-17}\,\,\text{e cm},\\
\text{Im}(d_\tau^{W})&< & 1{.}1 \times 10^{-17}\,\,\text {e cm},\label{wtau}
\end{eqnarray}
which again are far above the SM predictions $a_\tau^{W}=-(2{.}10+ 0{.}61 i)\times 10^{-6}$ \cite{Bernabeu:1994wh} and $d_\tau^{W}<8\times 10^{-34}$ e cm \cite{Booth:1993af}. However, several SM extensions predict large contributions to these observables that are closer to the experimental bounds. We will explore this possibility in the unparticle physics scenario proposed recently \cite{Georgi:2007ek}. In this framework, the gauge group is $SU(N)$ and there is a hidden sector in the low energy regime. It is conjectured that the theory remains conformal in the infrared (IR) regime, in such a way that there is a continuous mass spectrum. In this sense, the particle concept cannot be defined. Such a hidden sector would interact weakly with the SM via the exchange of heavy states and would manifest itself at an energy scale $\Lambda_{\mathcal U} > 1$ TeV. This scenario can have interesting consequences in both theoretical and phenomenological high energy physics. Despite the inherent complexity of the theoretical framework, the unparticle physics effects can be studied via an effective theory.  Indirect bounds on the unparticle parameters have been obtained from  LFV decays \cite{Moyotl:2011yv}, the muon anomalous magnetic moment \cite{Hektor:2008xu}, and monophoton production plus missing transverse energy at the LEP \cite{Cheung:2007ap}. More recently, experimental evidence of unparticles has
been searched for at the CERN LHC  \cite{Ask:2008fh}. In particular, the search for monojets plus missing transverse energy in  the 2010 LHC run data has allowed the CMS collaboration to impose strong bounds on the unparticle parameters \cite{Chatrchyan:2011nd}.

The physics of the $\tau$ lepton is expected to play an important role in the scientific program of present and future particle colliders \cite{Perl:1991gd,Gentile:1995ue,Pich:1997ym}. Because of its relatively large mass, the $\tau$ lepton can decay hadronically. From this class of processes, high precision measurements of several quantities can be extracted, such as the CKM matrix element $|V_{us}|$ and the mass of the strange quark. Also, as a result of its large variety of decay channels, the study of the $\tau$ lepton represents an interesting tool to search for CP violation, LFV, and other new physics effects. Although the ATLAS \cite{Aad:2011kt} and CMS \cite{Chatrchyan:2011nv} collaborations have already reported their first results for $\tau$ production from $Z$ decays, it is expected that the $B$ factories,  BABAR \cite{Aubert:2001tu} and  BELLE \cite{:2000cg}, collect large samples of data for $\tau^- \tau^+$ production. Furthermore, since these experiments use a center-of-mass energy around the $\Upsilon(4S)$ mass, they could be useful for the study of the electromagnetic properties of the $\tau$ lepton \cite{Bernabeu:2007rr,Inami:2002ur}.

The rest of the work is organized as follows. In Sec. II we present an overview of the effective  interactions of a spin-0 unparticle with a fermion pair. Section III is devoted to the analytical results for the fermion weak properties induced by a spin-0 unparticle, whereas the numerical results and discussion for the $\tau$ lepton is presented in Sec. IV, where a brief discussion on the bottom and top weak and electromagnetic properties is also included. The conclusions and outlook are presented in Sec. V.

\section{Unparticles Physics overview}

A toy model based on a scale invariant sector was already proposed some time ago by Banks and Zaks \cite{Banks:1981nn}, but it was only after the work of Georgi \cite{Georgi:2007ek} that high energy physicists  became more interested in this idea. Unparticle physics assumes the existence of a scale invariant hidden sector, known as
${\mathcal B}{\mathcal Z}$ sector, which can interact with the SM fields via the exchange of very heavy particles at a very high energy scale
${\mathcal M}_{\mathcal U}$. Below this energy scale, there are nonrenormalizable couplings between the fields of the ${\mathcal B}{\mathcal Z}$ sector and the SM ones. These couplings can be written generically as
$ {\mathcal O}_{SM}{\mathcal O}_{{\mathcal B}{\mathcal Z}}/{\mathcal M}_{\mathcal U}^{d_{SM}+d_{{\mathcal B}{\mathcal Z}}-4}$.  The dimension of the associated operators are $d_{{\mathcal B}{\mathcal Z}}$ and $d_{SM}$, respectively. Dimensional transmutation occurs at an energy scale $\Lambda_{\mathcal U}$ due to the renormalizable couplings of the ${\mathcal B}{\mathcal Z}$ sector. Below the scale $\Lambda_{\mathcal U}$, an effective theory can be used to describe the interactions between the fields of the ${\mathcal B}{\mathcal Z}$ sector and the SM fields, which arise from the exchange of unparticle fields. The corresponding effective Lagrangian that respects $SU_L(2)\times U_Y(1)$  gauge invariance can be written as \cite{Georgi:2007ek}:

\begin{equation}
{\mathcal L}_{\mathcal U}=C_{{\mathcal O}_{\mathcal U}} \frac{\Lambda_{\mathcal U}^{d_{{\mathcal B}{\mathcal Z}}-d_{\mathcal U}}}{{\mathcal M}_{\mathcal U}^{d_{SM}+d_{{\mathcal B}{\mathcal Z}}-4}} {\mathcal O}_{SM}{\mathcal O}_{\mathcal U},
\end{equation}
where $C_{{\mathcal O}_{\mathcal U}}$ is a coupling constant and the dimension, $d_{\mathcal U}$, of the unparticle operator, ${\mathcal O}_{\mathcal U}$, can be a fractionary number, though its value is restricted to the interval $1<d_\mathcal U<2$ due to unitarity  \cite{Georgi:2007ek,Grinstein:2008qk,Biggio:2008in,Nakayama:2007qu}. As far as the unparticle operators are concerned, their Lorentz structure can be constructed out of the operators ${\mathcal O}_{{\mathcal B}{\mathcal Z}}$ and their transmutation. In general there can be unparticle operators of scalar, ${\mathcal O}_{\mathcal U}$, vector, ${\mathcal O}_{\mathcal U}^\mu$, and tensor, ${\mathcal O}_{\mathcal U}^{\mu\nu}$, type. For simplicity we will only consider spin-0 unparticle operators in our analysis below. The effective Lagrangian describing the scalar and pseudo-scalar interactions of a spin-0 unparticle with a fermion pair is given by:

\begin{eqnarray}
{\mathcal L}_{{\mathcal U}^\text{spin-0}}&=&  \frac{\lambda_{ij}^{S}}{\Lambda_{\mathcal U}^{d_{\mathcal U}-1}} \bar{f}_i f_j {\mathcal O}_{\mathcal U} +\frac{\lambda_{ij}^{P}}{\Lambda_{\mathcal U}^{d_{\mathcal U}-1}} \bar{f}_i \gamma^5 f_j {\mathcal O}_{\mathcal U},\label{lup}
\end{eqnarray}
where $\lambda_{ij}^{{S,P}}=C_{{\mathcal O}_{\mathcal U}} \Lambda_{\mathcal U}^{d_{{\mathcal B}{\mathcal Z}}}/{\mathcal M}_{\mathcal U}^{d_{SM}+d_{{\mathcal B}{\mathcal Z}}-4}$ stands for the respective coupling constant.
Constraints on the coupling constant associated with the $\tau$ lepton have been already obtained from the LFV decay $\tau \to 3 \mu$ \cite{Moyotl:2011yv} and the muon anomalous magnetic moment \cite{Hektor:2008xu}. As far as the unparticle propagators are concerned, they are constructed using scale invariance and the spectral decomposition formula. The propagator for a spin-0 unparticle can be written as

 \begin{equation}
\Delta_F(p^2)= \frac{A_{d_\mathcal U}}{2\sin(d_\mathcal U \pi)} (-p^2-i\epsilon)^ {d_{\mathcal U}-2}, \label{unpro}
\end{equation}
where $A_{d_\mathcal U}$, which is meant to normalize the spectral density \cite{Cheung:2007ap}, is given by

\begin{equation}
A_{d_\mathcal U}=\frac{16\pi^2\sqrt{\pi}}{(2\pi)^{2d_\mathcal U}}\frac{\Gamma(d_\mathcal U+\frac{1}{2})}{\Gamma(d_\mathcal U-1)\Gamma(2d_\mathcal U)}.
\end{equation}
As expected, in the limit of $d_\mathcal U \to 1$, Eq.  (\ref{unpro}) becomes the propagator of a massless scalar particle.

 \section{Electromagnetic and weak properties of the fermions}

The electromagnetic and weak properties of fermions can be described through the following interaction Lagrangian:

\begin{eqnarray}
{\mathcal L}^{\text{spin}-1/2}&=&-\frac{i}{2} \bar{f} \sigma_{\mu\nu} \gamma_5 f(d_f F^{\mu\nu}_\gamma+d_f^{W} F^{\mu\nu}_Z)
                                            \nonumber\\
{}&{}&+\frac{e}{4m_f}\bar{f} \sigma_{\mu\nu} f(a_f F^{\mu\nu}_\gamma+a_f^{W} F^{\mu\nu}_Z),\label{lad}
\end{eqnarray}
where $F^{\mu\nu}_\gamma$ and $F^{\mu\nu}_Z$ are the electromagnetic and weak stress tensors, respectively. The fermion electromagnetic and weak properties arise at the loop level and can be extracted from the matrix element $ie\bar{\mathrm{u}}(p') \Gamma^\mu_V \mathrm{u}(p)$, where $ \Gamma_V^\mu$ is given by:

\begin{eqnarray}
\Gamma^\mu_V(q^2)&=& F_A(q^2)(\gamma^\mu \gamma_5 q^2-2{m_f} \gamma_5 q^\mu)+F_1(q^2) \gamma^\mu
                                            \nonumber\\
{}&{}&+F_2(q^2)i\sigma^{\mu\nu} q_\nu+F_3(q^2)\sigma^{\mu\nu} \gamma_5 q_\nu, \label{vad}
\end{eqnarray}
with $q=p'-p$ the four-momentum of the gauge boson $V$. The MDM and the EDM are given by $a_f=-2m_f F_2(q^2=0)$ and $d_f=-eF_3(q^2=0)$, whereas the weak properties, $a_f^{W}$ and $d_f^{W}$, are defined by analogue expressions but with the replacement  $q^2=m_Z^2$.

We now consider the flavor changing interaction given by Eq. (\ref{lup}) to obtain the WMDM and WEDM of the fermion $f$  induced by a spin-0 unparticle. We have calculated the loop amplitudes via Feynman parameters. The results can be written as
\begin{eqnarray}
a_f^{{W}}(d_\mathcal U)&=& \frac{ A_{d_\mathcal U} }{16 \pi^2\sin{(d_\mathcal U} \pi)} \sum_{i=e,\mu,\tau} \sqrt{r_{i}} \bigg( \frac{m_i^2}{\Lambda_{\mathcal U}^2} \bigg)^{d_\mathcal U-1} \int_0^1dx \int_0^{1-x}dy H(d_\mathcal U,r_{i},x_Z,x,y) \big[ g_V^f F_1(r_{i},x)+2 g_A^f F_2(r_{i},x)\big],\label{aw1}
\end{eqnarray}
and
\begin{eqnarray}
d_f^{{W}}(d_\mathcal U) &=&\frac{ -  e g_V^f A_{d_\mathcal U}}{16 \pi^2m_f\sin{(d_\mathcal U} \pi)} \sum_{i=e,\mu,\tau} \text{Im}\big( {\lambda_{f i}^P}^* \lambda_{f i}^S \big) \sqrt{r_i} \bigg( \frac{m_i^2}{\Lambda_{\mathcal U}^2} \bigg)^{d_\mathcal U-1}  \int_0^1dx  \int_0^{1-x}dy (1-x) H(d_\mathcal U,r_{i},x_Z,x,y),\label{dw1}
\end{eqnarray}
where we introduced the short-hand notation $r_{i}=m_f^2/m_i^2$,  $i$ stands for the flavor index of the internal fermion, and $g_{A,V}^f$ are the fermion coupling constants to the $Z$ gauge boson. We also introduced the following dimensionless functions:

\begin{eqnarray}
F_1(r_{i},x)&=& (x-1) \big(|\lambda_{f i}^S|^2(1+x\sqrt{r_{i}})+|\lambda_{f i}^P|^2(x\sqrt{r_{i}}-1) \big),\label{f1}
                                            \\
F_2(r_{i},x)&=& \sqrt{r_{i}}\,\text{Re}\big( {\lambda_{f i}^P}^* \lambda_{f i}^S \big)x(1-x), \label{f2}
                                          \\
H(d_\mathcal U,r_{i},x_Z,,x,y)&=&x^{1-d_\mathcal U}\left(r_{i}x_Z (x+y-1)y+(1-x)(1-r_{i}x)\right)^{d_\mathcal U-2},\label{h}
\end{eqnarray}
with $x_Z=m_Z^2/m_f^2$.
As expected, the fermion WEDM only receives contributions from the vector coupling $g_V^f$ and it is proportional to $\text{Im}\big( {\lambda_{f i}^P}^* \lambda_{f i}^S \big)$, which is expected as this property violates CP.  As a cross-check for our calculation, from Eqs. (\ref{aw1}) and (\ref{dw1}) we can obtain the fermion electromagnetic properties reported in Ref. \cite{Moyotl:2011yv} after the replacements $x_Z=0$, $g_A^f=0$ and $g_V^f=Q_f$ are done. Here $Q_f$ is the fermion electric charge in units of $e$. In the following section we will concentrate on the numerical evaluation of the electromagnetic and weak properties of the $\tau$ lepton, and also comment briefly on the respective properties of the bottom and top quarks.

\section{Numerical analysis and discussion}

The analysis of monophoton production plus missing transverse energy, $e^+e^-\to\gamma +X$, at the LEP was used in Ref. \cite{Cheung:2007ap} to impose a bound on the scale $\Lambda_{\mathcal U}$ as a function of $d_{\mathcal U}$. They considered the 95 \%C. L. limit  $\sigma(e^+e^-\to\gamma +X)\simeq 0{.}2$ pb obtained at $\sqrt{s}=207$ GeV by the L3 Collaboration. It was found that this limit requires $\Lambda_{\mathcal U}\ge 660$ TeV for $d_\mathcal U=1{.}4$ and $\Lambda_{\mathcal U}\ge 1{.}35$ TeV for  $d_\mathcal U= 2$. Stronger limits were obtained by the CMS collaboration using the data for monojet production plus missing transverse energy at the LHC for $\sqrt{s}=7$ TeV and an integrated luminosity of 35 pb$^{-1}$. Such data require  $\Lambda_{\mathcal U}\ge 10$ TeV for $d_\mathcal U=1{.}4$ and $\Lambda_{\mathcal U}\ge 1$ TeV for  $d_\mathcal U=1{.}7$ \cite{Chatrchyan:2011nd}. In summary, the region $d_\mathcal U< 1{.}4$ is strongly constrained as very large values of $\Lambda_{\mathcal U}$ are required. It is worth mentioning that in obtaining these bounds, the authors of Ref. \cite{Cheung:2007ap,Chatrchyan:2011nd} considered that the unparticle coupling constants have a magnitude of the order of unity. For our analysis we will consider  the intervals $1{.}7\leq d_\mathcal U\le 2$ for $\Lambda_{\mathcal U}=1$ TeV and $1{.}4\leq d_\mathcal U\le 2$ for $\Lambda_{\mathcal U}=10$ TeV.

To get an estimate of the electromagnetic and weak properties of the $\tau$ lepton, we will also need to make some assumptions concerning the magnitude of the coupling constants involved in the calculation. Based on our previous work \cite{Moyotl:2011yv}, we will consider the following hierarchy for the $\tau$ couplings $\lambda_{\tau e}^{S,P}<\lambda_{\tau \mu}^{S,P}\ll\lambda_{\tau \tau}^{S,P}$. It means that we will assume that LFV interactions occur mainly between the $\mu$ and $\tau$ leptons. Therefore, we will neglect the contributions from the $\lambda_{\tau e}^{S,P}$ coupling. In addition, for the flavor conserving couplings we will assume that $\lambda_{\mu \mu}^{S,P}\simeq\lambda_{\tau \tau}^{S,P}$. As far as the numerical values of the coupling constants are concerned, we will consider values that are consistent with the bounds obtained in Ref. \cite{Moyotl:2011yv} from the experimental limits on the muon MDM and the LFV decay $\tau\to3\mu$. In particular we will consider the value  $\lambda_{\tau \tau}^{S,P}\simeq 1{.}6$ (1.0), which is consistent with $d_\mathcal U=1{.}7$ (1.4), and  $\Lambda_{\mathcal U}=1$ TeV (10 TeV). Also, when analyzing the MDM and WMDN we will consider the following scenarios:

\begin{itemize}
\item Lone contribution  from the scalar coupling: $\lambda_{\tau i}^P=0$ and $\lambda_{\tau i}^S\ne0$.
\item Lone contribution from the pseudo-scalar coupling: $\lambda_{\tau i}^P\ne0$ and $\lambda_{\tau i}^S=0$.
\item Contribution from both  scalar and pseudo-scalar couplings: $\lambda_{\tau i}^S\simeq \lambda_{\tau i}^P$.
\end{itemize}
In the case of the EDM and WEDM, since they require the simultaneous contribution from both scalar and pseudo-scalar couplings, we will only consider the last scenario.

\subsection{$\tau$ magnetic dipole moment}

\begin{figure}[!hbt]
\centering
\includegraphics[width=8.5cm]{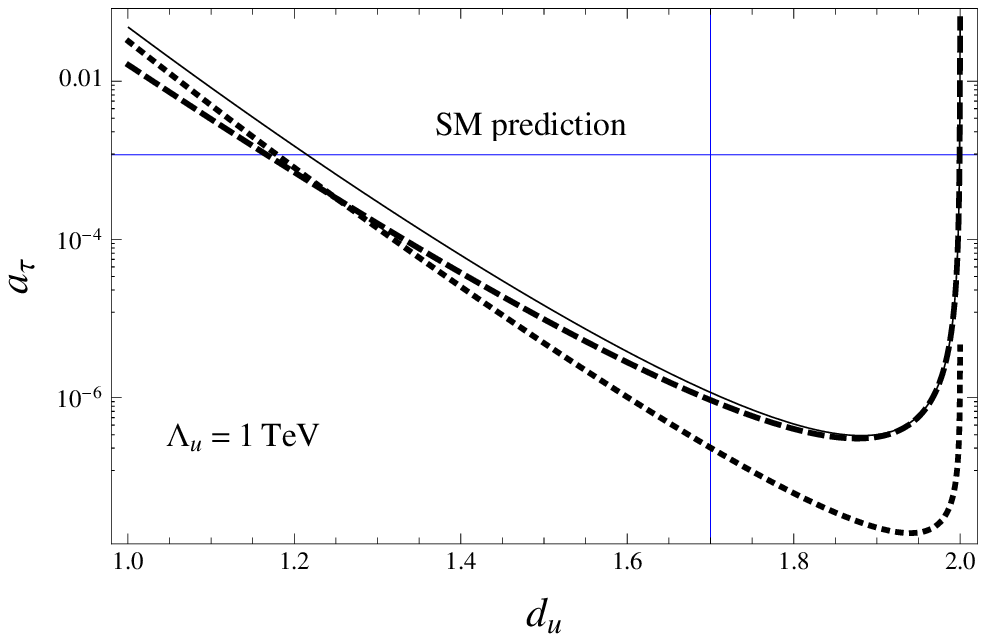}
\includegraphics[width=8.5cm]{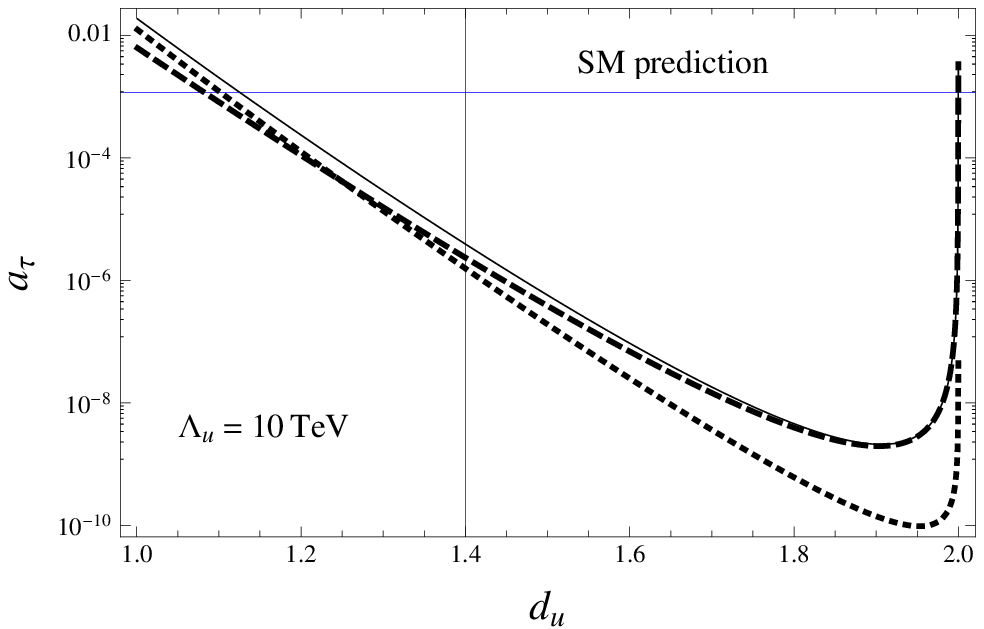}
\caption{Contribution from a spin-0 unparticle to the $\tau$ MDM as a function of the scale dimension
$d_\mathcal U$ for two values of $\Lambda_\mathcal U$.  We show the pure scalar contribution (solid line), the absolute value of the pure pseudo-scalar contribution (dashed line), and the total contribution (dotted line). The horizontal line is the SM contribution, and the vertical line represents the lower bound obtained by the CMS collaboration \cite{Chatrchyan:2011nd}.}
\label{aem1-10}
\end{figure}

Numerical evaluation of the $\tau$ MDM induced by a spin-0 unparticle shows that  the pseudo-scalar contribution is negative whereas the scalar contribution is positive, with the latter slightly larger in magnitude than the former. We have plotted in Fig. \ref{aem1-10} the  scalar contribution, the absolute value of the pseudo-scalar contribution, and the total contribution for
$\Lambda_{\mathcal U}=1$ TeV  and $\Lambda_{\mathcal U}=10$ TeV. We also included the SM prediction, which is given by  the horizontal  line. Since the unparticle propagator contains the term $\sin(d_\mathcal U\pi)$ in the denominator,  the contributions to the MDM diverge when $d_\mathcal U \to 2$. Therefore, in the allowed area, the largest contributions to the $\tau$ MDM are reached for $d_\mathcal U$ around 2. In this case $a_\tau^{\mathcal U}$ can be of the order of the SM contribution or larger. On the other hand, for values of $d_\mathcal U$ close to the lower bound, $a_\tau^{\mathcal U}$  is of the order of $10^{-6}$. When $\Lambda_{\mathcal U}=10$ TeV, the unparticle contribution to the $\tau$ MDM is more suppressed and its lowest values are of the order of $10^{-9}-10^{-10}$ . Since the scalar and pseudo-scalar contributions to $a^{\mathcal U}_\tau$ are about the same order of magnitude but opposite in sign, the total contribution can cancel out largely, which is more evident for $d_\mathcal U$ around 1.9. Therefore the largest contribution to $a^{\mathcal U}_\tau$ would arise in the scenario when only one contribution,
scalar or pseudoscalar, is present and for low values of  $\Lambda_\mathcal U$. For $d_\mathcal U$ around
$1{.}95$, both scalar and pseudo-scalar contributions reach their minimal absolute values: $|a_\tau^\mathcal U|\simeq 5\times10^{-7}$ when $\Lambda_{\mathcal U}=1$ TeV and $|a_\tau^\mathcal U|\simeq3\times10^{-9}$ when $\Lambda_{\mathcal U}=10$ TeV.

It is interesting to make a comparison between our results and those arising in other SM extensions, such as the SeeSaw model and an extension of the MSSM with a mirror fourth generation. While the type-I and type-III SeeSaw models predict the contributions $|a_\tau^{I}|<1{.}87\times10^{-8}$ and $|a_\tau^{III}|<7{.}55\times10^{-9}$ \cite{Biggio:2008in}, for representative values of the model parameters, the extension of the MSSM with a mirror fourth generation predicts a positive contribution, of the order of $10^{-6}-10^{-9}$ \cite{Ibrahim:2008gg}. The $\tau$ MDM has also been studied in the framework of  the  effective Lagrangian approach and the Fritzsch-Xing lepton mass matrix, but the respective contributions are even more suppressed \cite{adtau-eff}, of the order of $10^{-11}$. We thus conclude that the contribution from a spin-0 unparticle to the $\tau$ MDM can be of the same order of magnitude and even larger than the predictions of other SM extensions.

\begin{figure}[!hbt]
\centering
\includegraphics[width=8.5cm]{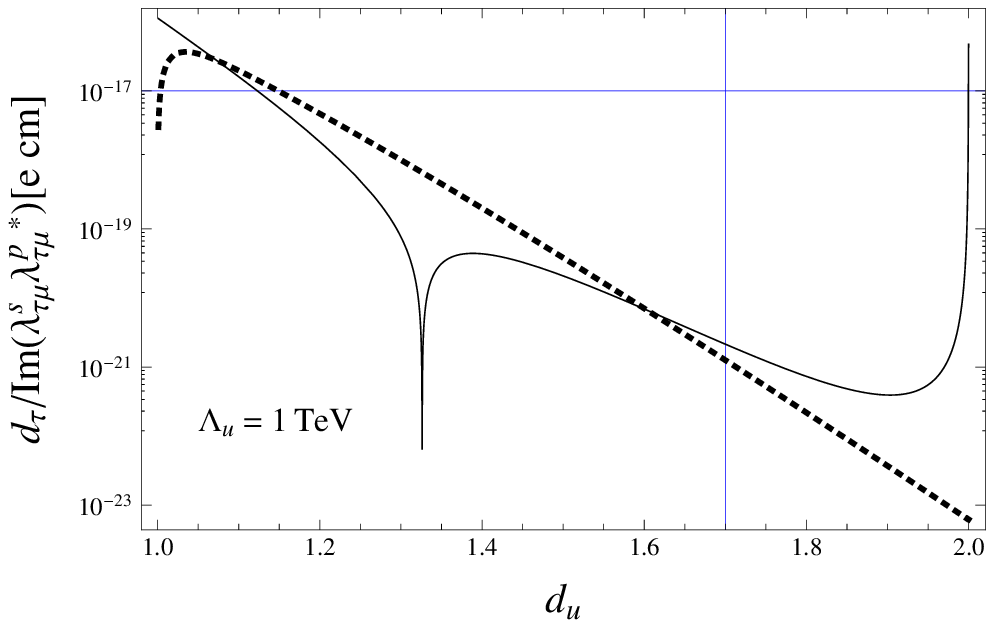}
\includegraphics[width=8.5cm]{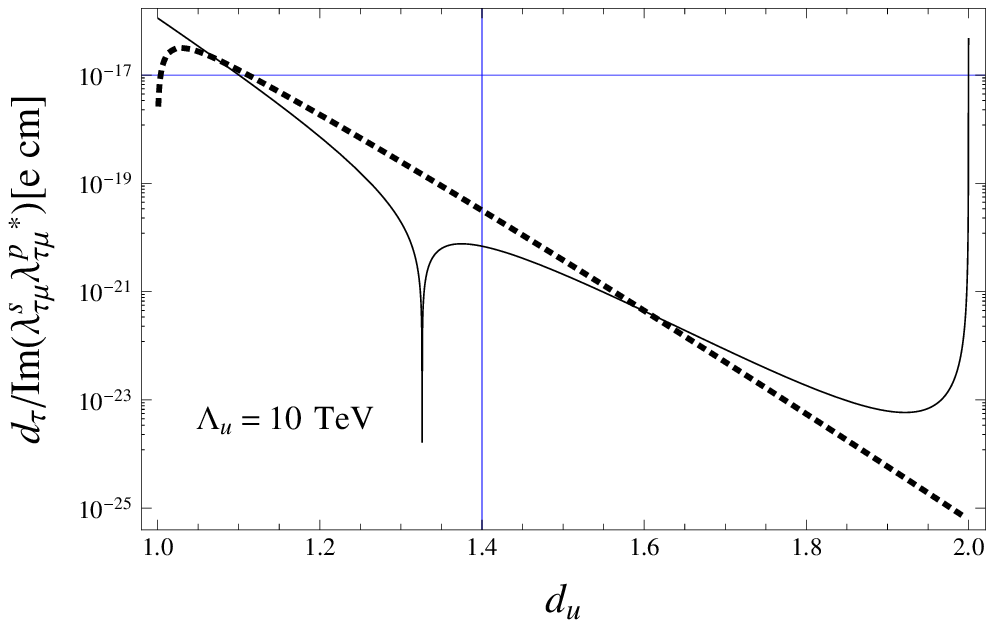}
\caption{Absolute values of the real (solid line) and imaginary (dotted line) parts of the contribution from a spin-0 unparticle to the $\tau$ EDM as a function of the scale dimension
$d_\mathcal U$ for two values of $\Lambda_\mathcal U$. The horizontal line is the SM contribution, and the vertical line represents the lower bound obtained by the CMS collaboration \cite{Chatrchyan:2011nd}.}
\label{dem1-10}
\end{figure}

\subsection{$\tau$ electric dipole moment}
The EDM requires an internal fermion in the loop different than the external fermion, so it must be induced by flavor changing couplings. Furthermore, it is necessary the presence of a CP-violating phase in the constant couplings. We thus write
$\text{Im}\big( \lambda_{\tau\mu}^{P*} \lambda_{\tau\mu}^S \big)=|\lambda_{\tau\mu}^S||\lambda_{\tau\mu}^P| \sin\Delta \Phi_{\tau\mu}^{S,P}$, where $\Delta\Phi_{\tau\mu}^{S,P}=\theta_{\tau\mu}^S-\theta_{\tau\mu}^P$ is the relative phase between the scalar and pseudo-scalar couplings. It is only the relative CP-violating phase that must be nonzero in order to have an EDM. Depending on  this phase, the EDM can be negative or positive, which poses no problem as the experimental bound also comprehends negative values. In order to analyze the unparticle contribution to the $\tau$ EDM, we will not  consider specific values for $\text{Im}\big( \lambda_{\tau\mu}^{P*} \lambda_{\tau\mu}^S \big)$. In Fig.  \ref{dem1-10} we have plotted the absolute values of the real and imaginary parts of the contribution to the $\tau$ EDM from a spin-0 unparticle as a function of the scale $d_\mathcal U$ for $\Lambda_{\mathcal U}=1$ TeV  and $\Lambda_{\mathcal U}=10$ TeV. A detailed analysis allows us to conclude that  there is a change in the sign of the real part of the $\tau$ EDM at $d_\mathcal U\simeq 1{.}325$, whereas its imaginary part is always positive. In the allowed region, both the real and imaginary parts are positive, although the CP-violating phase can give an additional change of sign. It is also interesting that the real part diverges when $d_\mathcal U \to 2$, but the imaginary part is negligibly small. Therefore, around  $d_\mathcal U=2$, the $\tau$ EDM is almost real and also reaches its largest size. At the lowest allowed value of $d_\mathcal U$, the contributions to the $\tau$ EDM are $d_\tau^{\mathcal U}=\text{Im}\big( \lambda_{\tau\mu}^{P*} \lambda_{\tau\mu}^S \big) (2{.}14+1{.}25 i) \times10^{-21}$ e cm when $\Lambda_\mathcal U=1$ TeV and $d_\tau^{\mathcal U}=\text{Im}\big( \lambda_{\tau\mu}^{P*} \lambda_{\tau\mu}^S \big) (0{.}69+3{.}17 i) \times10^{-20}$ e cm when $\Lambda_{\mathcal U}=10$ TeV. It  can also be observed that, in the allowed region, the real part reaches its minimal value at $d_\mathcal U\simeq 1{.}9$, where  $d_\tau^{\mathcal U}=\text{Im}\big( \lambda_{\tau\mu}^{P*} \lambda_{\tau\mu}^S \big) (3{.}92+0{.}39 i) \times10^{-22}$ e cm  when  $\Lambda_{\mathcal U}=1$ TeV, and $d_\tau^{\mathcal U}=\text{Im}\big( \lambda_{\tau\mu}^{P*} \lambda_{\tau\mu}^S \big) (5{.}85+0{.}40 i) \times10^{-24}$ e cm when $\Lambda_{\mathcal U}=10$ TeV.  In general, the spin-0 unparticle contribution to $d_\tau$ can be above the SM prediction \cite{dsm} as long as $\text{Im}\big( \lambda_{\tau\mu}^{P*} \lambda_{\tau\mu}^S \big)$ is not too small.

As far as other SM extensions are concerned, in an extension of the MSSM with vectorlike multiplets, the contributions to the $\tau$ EDM arise at the one loop level from loops carrying  $W$ gauge bosons, charginos (${\tilde{\chi}}_i^\pm$) or neutralinos (${\tilde{\chi}}_i^0$). Since these particles are heavier than the $\tau$ lepton, their contributions to the EDM are purely real and  have values ranging from $d_\tau\simeq 6{.}5\times10^{-18}$ e cm to $d_\tau\simeq 3{.}0\times10^{-23} $ e cm \cite{Ibrahim:2010va}. In contrast, the unparticle contribution $d_\tau^{\mathcal U}$ is almost real at $d_\mathcal U\simeq 2$, where it can reach values of the order of  $10^{-18}$ e cm, though it tends to be smaller for other $d_\mathcal U$ values.  The $\tau$ EDM has also been studied in other SM extensions, but the respective predictions  were found to be very small. This is the case of the framework of the Fritzsch-Xing lepton mass matrix, in which $|d_\tau|<2{.}2\times 10^{-25}$ e cm \cite{adtau-eff}.

\begin{figure}[!hbt]
\centering
\includegraphics[width=8.5cm]{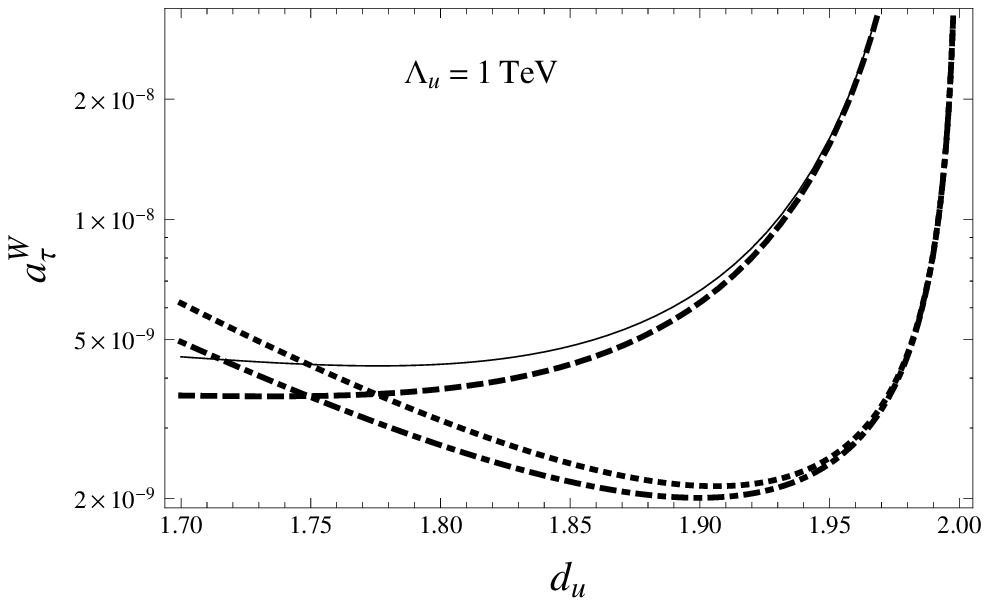}
\includegraphics[width=8.5cm]{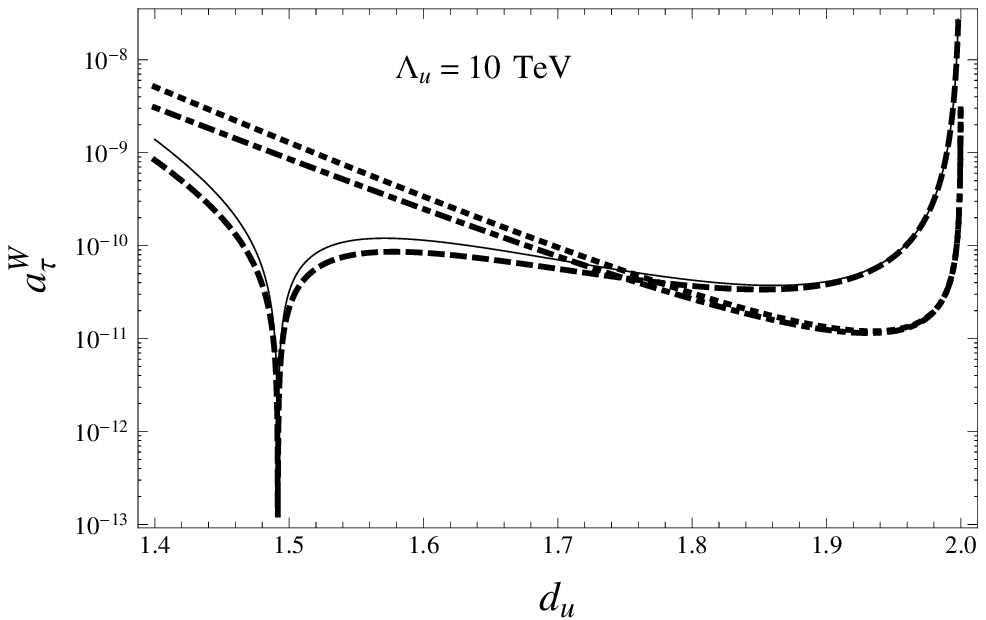}
\caption{Absolute values of the real and imaginary parts of the contribution from scalar (solid and dotted lines, respectively) and pseudo-scalar (dashed and dot-dashed lines, respectively) unparticle couplings to the $\tau$ WMDM, as a function of the scale dimension $d_\mathcal U$ and for two values of $\Lambda_\mathcal U$. The range of the horizontal axis corresponds to the region still allowed according to CMS \cite{Chatrchyan:2011nd}.}
\label{aw1-10}
\end{figure}

\subsection{$\tau$ weak magnetic dipole moment}

Contrary to the case of the MDM, the WMDM does depend on the relative phase
$\Delta\Phi_{\tau i}^{S,P}$, though a CP violating phase is not required. Furthermore, since such a phase only appears when there is flavor changing couplings, which can be neglected as compared to the flavor conserving ones, for simplicity we will consider a vanishing $\Delta\Phi_{\tau\tau}^{S,P}$. We will first examine the individual behavior of the scalar and pseudo-scalar contributions. To this end, we show in Fig. \ref{aw1-10} the absolute values of the real  and imaginary parts  of both the scalar  and pseudo-scalar contributions to the $\tau$ WMDM for $\Lambda_{\mathcal U}=1$ TeV  and $\Lambda_{\mathcal U}=10$ TeV. In this Figure we can observe that both contributions show a similar behavior although the magnitude of the scalar contribution is slightly larger. Moreover, the largest values of both contributions can arise around $d_\mathcal U \to 2$, similar to what is observed in the MDM. It is also interesting to note that for
$\Lambda_{\mathcal U}=1$ TeV, the scalar contributions are negative in the whole $d_\mathcal U$ interval, while the pseudo-scalar contributions are positive, contrary to the case of the MDM.  For the same value of $\Lambda_\mathcal U$, we also observe that the magnitude of the $\tau$ WDM can be of the order of $10^{-9}$ in the allowed region of $d_\mathcal U$. The situation changes when  $\Lambda_{\mathcal U}=10$ TeV,  in which case  the real part of the scalar contribution changes sign from positive to negative at $d_\mathcal U \simeq 1{.}49$, whereas the imaginary part remains positive. The real and imaginary parts of the pseudo-scalar contribution also show a similar behavior but they are of opposite sign to their scalar analogues. Thus, when $\Lambda_\mathcal U=10$ TeV, the $\tau$ WDM is purely imaginary at $d_\mathcal U\simeq 1.49$, with a magnitude of the order of $10^{-9}$, whereas for other $d_\mathcal U$ values, the magnitude of the real and imaginary parts fall in the range $10^{-8}-10^{-11}$.

\begin{figure}[!hbt]
\centering
\includegraphics[width=8.5cm]{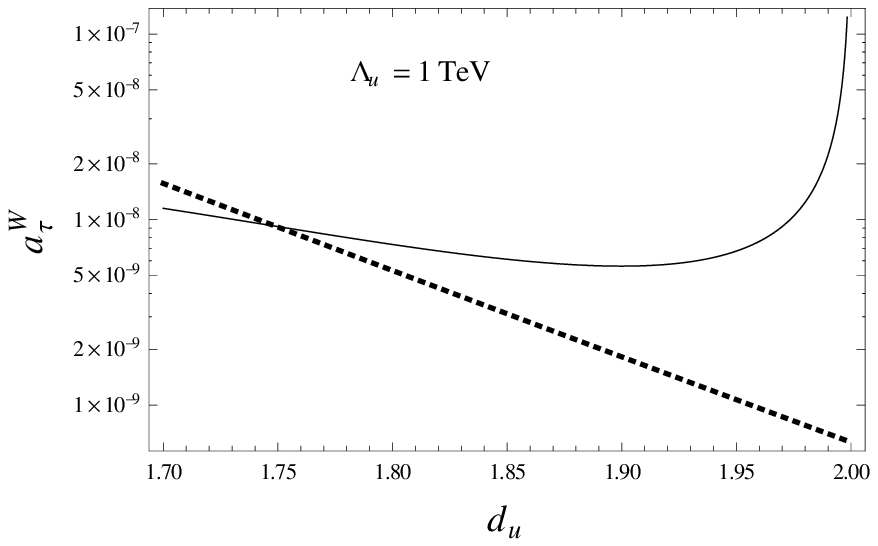}
\includegraphics[width=8.5cm]{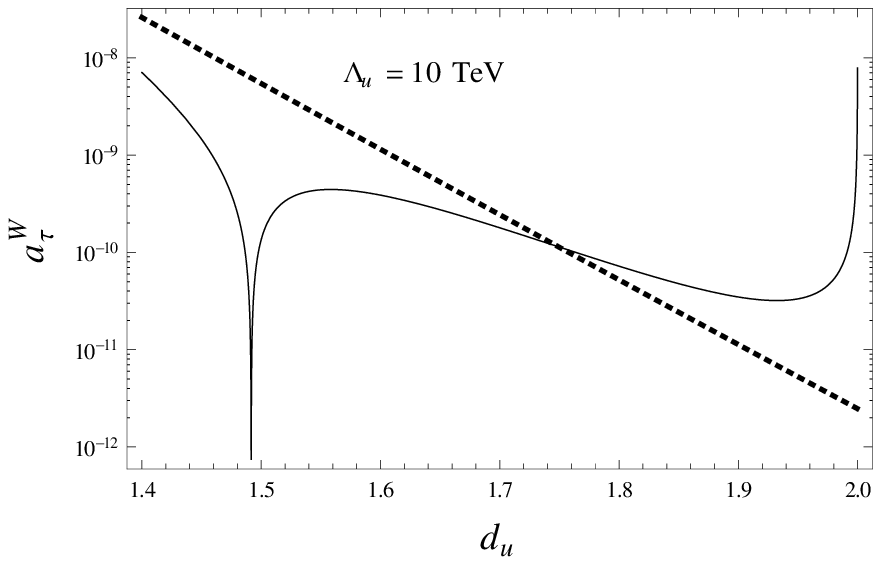}
\caption{Absolute values of the real (solid line) and imaginary (dotted line) parts of the total contribution of a spin-0 unparticle to the $\tau$ WMDM as a function of the scale $d_\mathcal U$ for two values of $\Lambda_\mathcal U$. For simplicity we neglect LFV and use $\Delta\Phi_{\tau\tau}^{S,P}=0$.  The range of the horizontal axis corresponds to the region still allowed according to CMS \cite{Chatrchyan:2011nd}.}
\label{awphi}
\end{figure}

As mentioned above, the LFV contributions to the $\tau$ WMDM are expected to be subdominant. Since the relative phases of the flavor conserving couplings are zero, we have $\text{Re}\big( {\lambda_{\tau \tau}^P}^* \lambda_{\tau \tau}^S \big)=|\lambda_{\tau \tau}^P| |\lambda_{\tau \tau}^S|$. Therefore, apart from the individual contributions from scalar and pseudo-scalar couplings, there are an interference term that contributes to the WMDM. In Fig. \ref{awphi} we show the absolute values of the real and imaginary parts of the total contribution of a spin-0 unparticle to the $\tau$ WMDM for $\Lambda_{\mathcal U}=1$ TeV and $\Lambda_{\mathcal U}=10$ TeV. We note that while the real part diverges when $d_\mathcal U \to 2$, the imaginary part almost vanishes, which is similar to the behavior of the EDM, as shown in Fig. \ref{dem1-10}. It is also observed that when  $\Lambda_{\mathcal U}=1$ TeV, both the real and imaginary parts are positive, but when $\Lambda_{\mathcal U}=10$ TeV only the imaginary part is positive whereas the real part changes from negative to positive at $d_\mathcal U\simeq 1.49$. At the lowest allowed value of $d_\mathcal U$,  $a_\tau^{W}\simeq (1{.}15+i1{.}56)\times10^{-8}$ when $\Lambda_\mathcal U=1$ TeV and $a_\tau^{W}\simeq (-0{.}70+i 2{.}59\times)10^{-8}$ when $\Lambda_\mathcal U=10$ TeV. The minimal values of the real contribution to $a_\tau^\mathcal U$  are reached at $d_\mathcal U$ around  $1.9$, and they correspond to $a_\tau^{W}\simeq (5{.}61+i1{.}82)\times10^{-9}$  when $\Lambda_\mathcal U=1$ TeV and $a_\tau^{W}\simeq (3{.}19+i 0{.}71\times)10^{-11}$ when $\Lambda_\mathcal U=10$ TeV.

Finally, we would like to compare our predictions with  the  SM prediction \cite{Bernabeu:1994wh}. In the most promising scenario, the unparticle contributions are about two orders of magnitude smaller than the SM contribution, but in a more conservative scenario they are about five orders of magnitude below. The unparticle scenario however allows for both positive or negative contributions.

\subsection{$\tau$ weak electric dipole moment}

\begin{figure}[!hbt]
\centering
\includegraphics[width=8.5cm]{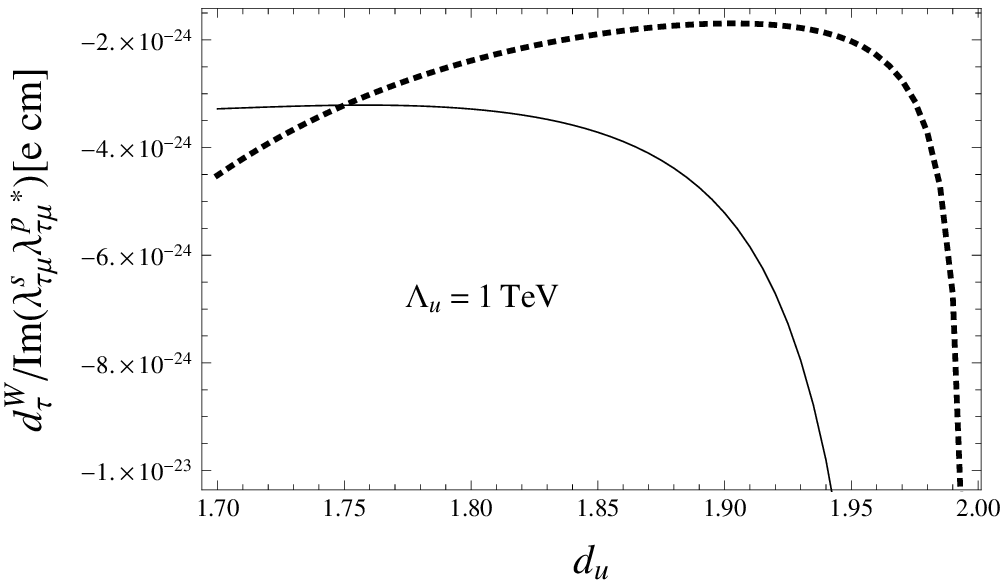}
\includegraphics[width=8.5cm]{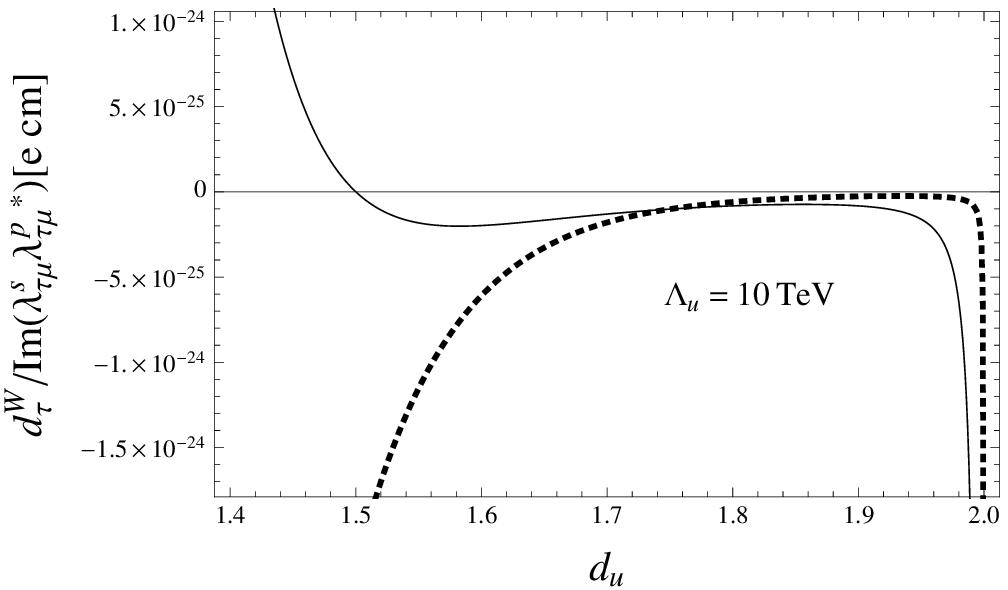}
\caption{The same as in Fig \ref{dem1-10}, but for the $\tau$ WEDM.}
\label{dwt}
\end{figure}

As was the case with the EDM, a nonzero WEDM requires a CP-violating phase, which appears when LFV couplings are present. Thus, in order to analyze this property we will follow the same approach used to calculate the EDM. However, before the numerical evaluation, the analysis of Eq. (\ref{h}) suggests  that the $\tau$ WEDM is expected to be smaller than the EDM due to the term proportional to
$x_Z$. We show in Fig.  \ref{dwt} the behavior of the real and imaginary parts of the $\tau$ WEDM induced by a spin-0 unparticle as a function of $d_\mathcal U$ and for two values of $\Lambda_\mathcal U$.  As was anticipated, this property shows a behavior similar to that of the $\tau$ EDM, though it has a smaller magnitude and opposite sign. When $\Lambda_{\mathcal U}=1$ TeV, both the real and imaginary contributions are negative, but when $\Lambda_{\mathcal U}=10$ TeV, the imaginary part is negative whereas the real part changes from positive to negative at $d_\mathcal U \simeq 1{.}5$, where the $\tau$ WEDM is almost imaginary, i.e. $d_\tau^{W}\simeq-i\text{Im}\big( \lambda_{\tau\mu}^{P*} \lambda_{\tau\mu}^S \big)2{.}2\times10^{-24}$ e cm. Contrary to behavior of the EDM, both  the real and imaginary parts of the WEDM diverge when $d_\mathcal U\to 2$. Also, at the lowest allowed value of $d_\mathcal U$,
$d_\tau^{W}\simeq -\text{Im}\big( \lambda_{\tau\mu}^{P*} \lambda_{\tau\mu}^S \big)(3{.}28 + i4{.}51)\times10^{-24}$ e cm when $\Lambda_{\mathcal U}=1$ TeV and   $d_\tau^{W}\simeq\text{Im}\big( \lambda_{\tau\mu}^{P*} \lambda_{\tau\mu}^S \big)(2{.}67-i8{.}24)\times10^{-24}$ e cm when $\Lambda_{\mathcal U}=10$ TeV. When
$\Lambda_{\mathcal U}=1$ TeV, the minimal value of the real part is  reached at $d_\mathcal U \simeq 1{.}76$, where $d_\tau^{W}=-\text{Im}\big( \lambda_{\tau\mu}^{P*} \lambda_{\tau\mu}^S \big)(3{.}21 + i3{.}01)\times10^{-24}$ e cm, while the minimal value of the imaginary part is reached at $d_\mathcal U \simeq 1{.}9$, where $d_\tau^{W}= -\text{Im}\big( \lambda_{\tau\mu}^{P*} \lambda_{\tau\mu}^S \big)(5{.}21 + i1{.}69)\times10^{-24}$ e cm. On the other hand, when $\Lambda_{\mathcal U}=10$ TeV the minimal values of the real and imaginary parts are reached at $d_\mathcal U \simeq 1{.}86$ and  $d_\mathcal U = 1{.}93$, respectively. These minimal values correspond to  $d_\tau^{W}= -\text{Im}\big( \lambda_{\tau\mu}^{P*} \lambda_{\tau\mu}^S \big)(7{.}40 + i3{.}48)\times10^{-26}$ e cm and $d_\tau^{W}=-\text{Im}\big( \lambda_{\tau\mu}^{P*} \lambda_{\tau\mu}^S \big)(1{.}09 + i0{.}24)\times10^{-25}$ e cm, respectively.

Although the contribution from a spin-0 unparticle to the $\tau$ WEDM can be up to ten orders of magnitude larger than the SM contribution in the best scenario, there is a difference of about seven orders of magnitude with respect to the experimental limits, although the actual value of $\text{Im}\big( \lambda_{\tau\mu}^{P*} \lambda_{\tau\mu}^S \big)$ can reduce additionally  the size of the unparticle contribution or  change its sign.  In summary, the unparticle scenario allows for negative or positive contributions to the $\tau$ WEDM, which can be closer to the experimental limits than the SM contribution.

\subsection{Electromagnetic and weak properties of heavy quarks }

Due to quark confinement, the study of the quark properties requires more elaborated experimental techniques. Indirect measurements can be extracted from composite states such as the neutron, the proton, the deuteron, or atoms of thallium and ${}^{199}$Hg. For instance, the MDM of the charm and bottom quarks can be extracted from the heavy baryons  $\Sigma_c$, $\Lambda_c$, $\Xi_c$ and $\Xi_b$\cite{dm-hb}. On the other hand, it has been proposed that the top quark properties could be measured at hadron colliders via  $ V t\bar{t}$ production, with the top quarks decaying in the dominant channel $t\to W b$. Along these lines, the authors of Ref.
\cite{Baur:2004uw} have shown that the LHC would allow one to measure anomalous contributions to the $\gamma t \bar{t}$ coupling via the process $pp\to \gamma \bar{t} t\to \gamma W^+ \bar{b} W^- b$ as long as one of the $W$ gauge bosons decays leptonically $W\to l \nu_l$ and the other one decays hadronically $W\to jj$. When $V=Z$, due to the trigger efficiency, it is assumed that the $Z$ gauge boson decays leptonically $Z\to \bar{l'} l'$, with either one of the $W$ gauge boson decaying leptonically and the other one decaying hadronically or with both of them decaying hadronically. In this scenario, it was shown that the measurements at the LHC  would be sensitive enough to allow one to extract anomalous contributions to the $Z t \bar{t}$ couplings  from the data on the processes $p p \to l'\bar{l'} l \nu_l b \bar{b} jj$ and $p p \to l'\bar{l'} b\bar{b}+4j$, as long as the luminosity is increased by a 10 factor (SLHC). A more detailed discussion on the technical details of this analysis is beyond the present work, so we refer the reader to the original references. We will content ourselves with analyzing the potential unparticle effects on the electromagnetic and weak properties of heavy quarks, namely the bottom quark and the top quark. We will consider the scenarios discussed above for the unparticle parameters and the coupling constants. Since the behavior of the electromagnetic and weak properties of heavy quarks is similar to the one observed in the $\tau$ lepton case, we will only present an estimate of the unparticle contributions  at the lowest allowed value of $d_\mathcal U$. Also, we will present the minimal values of the real part, which are obtained for $d_\mathcal U$ around 1.9. We show the results in Tables \ref{ew-b} for the bottom quark and Table \ref{ew-t} for the top quark.

\begin{table}[!htb]
\begin{center}
\begin{tabular}{|c|cccc|}
\hline
{}	& $\Lambda_{\mathcal U}=1$ TeV, ${d_\mathcal U = 1{.}7}$	&  $\Lambda_{\mathcal U}=10$ TeV, ${d_\mathcal U = 1{.}4}$ & $\Lambda_{\mathcal U}=1$ TeV	 &$\Lambda_{\mathcal U}=10$ TeV	\\
\hline
\hline
$a_b$	&	$2{.}55\times10^{-8}$ &	$1{.}17\times10^{-7}$ &	$2{.}80\times10^{-9}$&	 $1{.}40\times{10}^{-11}$\\
$d_b/\text{Im}({\lambda_{bs}^P}^*\lambda_{bs}^S)$	&	 $(1{.}63+i1{.}55)\times10^{-23}$ &	$(-0{.}11+i1{.}07)\times10^{-22}$&	 $(2{.}99+i0{.}39)\times10^{-24}$ & $(1{.}69+i0{.}22)\times10^{-26}$ \\
$a_b^{W}$&	$(0{.}73+i1{.}00)\times10^{-8}$ &	$(-0{.}42+i1{.}66)\times10^{-8}$&	 $(3{.}59+i1{.}16)\times10^{-9}$ &	$(2{.}05+i0{.}49)\times10^{-11}$\\
$d_b^{W}/\text{Im}({\lambda_{bs}^P}^*\lambda_{bs}^S)$&	 $(-4{.}34-i5{.}98)\times10^{-26}$ &	$(1{.}37-i4{.}23)\times10^{-26}$&	 $(-6{.}91-i2{.}24)\times10^{-26}$&	$(-6{.}69-i1{.}27)\times10^{-28}$\\
\hline
\end{tabular}
\caption{Contributions from a spin-0 unparticle to the electromagnetic
and weak properties of the bottom quark. The values shown
are those obtained at the lowest allowed value of $d\uu$ (second and third
columns) together with the values that correspond to the minimal value
of the real part (fourth and fifth columns) for two values of $\Lambda_\uu$.}
\label{ew-b}
\end{center}
\end{table}

Unfortunately,  there are no other theoretical predictions in the case of the bottom quark to our knowledge. However, due to its heavy mass, it has been suggested that the top quark may be sensible to new physics effects. This has motivated the study of the top quark properties, such as the EDM \cite{Ibrahim:2010hv}, WEDM \cite{Hollik:1998vz}, the chromoelectric dipole moment  \cite{Ibrahim:2011im}, and the chromomagnetic dipole moment \cite{Martinez:2008hm}. Despite its heaviness, the top EDM is predicted to be negligibly small in the SM, i.e. $d_t<10^{-30}$ e cm \cite{Soni:1992tn}, which is of the same order of magnitude than that of the $\tau$ lepton. As far as other extensions of the SM are concerned, an MSSM extension with vectorlike multiplets predicts values for $d_t$ ranging from $2{.}87\times10^{-19}$ e cm to $2{.}85\times10^{-22}$ e cm \cite{Ibrahim:2010hv}.
The unparticle contribution $d_t^{\mathcal U}$ has an imaginary part due to the internal charm quark but its order of magnitude is below the $10^{-19}$ level, although at best it can be of the same order of magnitude than the contributions predicted in the MSSM extension with vectorlike multiplets. As for the top WEDM, it was calculated long ago in the framework of the R-parity preserving MSSM version with complex parameters, where it was found that $d_t^{W}\simeq(0{.}351-1{.}264)\times10^{-19}$ \cite{Hollik:1998vz}. However, the order of magnitude of this prediction can have a significant decrease if updated bound on the model parameters are considered. On the other hand, the unparticle contribution is much smaller and can be up to five orders of magnitude below.

\begin{table}[!htb]
\begin{center}
\begin{tabular}{|c|cccc|}
\hline
{}	& $\Lambda_{\mathcal U}=1$ TeV, ${d_\mathcal U = 1{.}7}$	&  $\Lambda_{\mathcal U}=10$ TeV, ${d_\mathcal U = 1{.}4}$ & $\Lambda_{\mathcal U}=1$ TeV	 &$\Lambda_{\mathcal U}=10$ TeV	\\
\hline
$a_t$	&	$-2{.}05\times10^{-5}$ &	$-7{.}21\times10^{-6}$ &	 $-1{.}32\times10^{-5}$&	$-8{.}52\times{10}^{-8}$\\
$d_t/\text{Im}({\lambda_{tc}^P}^*\lambda_{tc}^S)$ & $(-2{.}42-i2{.}58)\times10^{-20}$ &	$(0{.}23-i1{.}32)\times10^{-20}$  & $(-2{.}10-i0{.}72)\times10^{-20}$  & $(-1{.}47-i0{.}32)\times10^{-22}$ \\
$a_t^{W}$ &	$-4{.}73\times10^{-5}$ &	$-2{.}06\times10^{-5}$&	 $-2{.}77\times10^{-5}$ &	$-1{.}74\times10^{-7}$\\
$d_t^{W}/\text{Im}({\lambda_{tc}^P}^*\lambda_{tc}^S)$& $(4{.}50+i6{.}19)\times10^{-22}$	&$(-0{.}63+i1{.}94)\times10^{-22}$ &$(9{.}34+i3{.}03)\times10^{-22}$&	$(6{.}88+i1{.}76)\times10^{-24}$\\
\hline
\end{tabular}
\caption{The same as in Table \ref{ew-b}, but for the top quark. }
\label{ew-t}
\end{center}
\end{table}

\section{Conclusions}

We have studied the contribution to the electromagnetic and weak properties of fermions from a spin-0 unparticle, with particular emphasis on the $\tau$ lepton properties.  As far as the unparticle parameters are concerned, we considered the most recent CMS bounds from monojet production plus missing transverse energy at the LHC, while for the coupling constants we used the indirect limits obtained from the experimental bounds on  LFV decays and the muon MDM. In the most promising scenario, we find that the unparticle contribution to the electromagnetic properties of the $\tau$ lepton can be larger than the contributions predicted by the SM and some of its extensions, such as the SeeSaw model and  extensions of the minimal supersymmetric standard model with a mirror fourth generation and vectorlike multiplets.  As far the $\tau$ weak properties  are concerned, the contributions from a spin-0 unparticle are smaller than the respective contributions to  the electromagnetic properties, although they are larger than the SM contributions, tough much smaller than the current experimental limits. We also examine the electromagnetic and weak properties of the bottom and top quarks. In particular, we find that the predictions obtained for the top EDM in the framework of unparticle physics are of  similar order of magnitude than in an MSSM extension with vectorlike multiplets. In general, the most promising scenario for the contribution of unparticle physics to the electromagnetic and weak properties of fermions is that in which  $d_\mathcal U$ is close to 2, which is a region still allowed by the most recent constraints on unparticle physics from the LHC data. We would like to emphasize however that our results depend considerably on the values of the scale $\Lambda_\mathcal U$ and the dimension $d_\mathcal U$.

\acknowledgments{We acknowledge financial support from Conacyt and SNI (M\'exico). G.T.V also acknowledges VIEP-BUAP for partial support.}


\begin{thebibliography}{9}
\bibitem{Kobayashi:1973fv}
  M.~Kobayashi and T.~Maskawa,
  Prog.\ Theor.\ Phys.\  {\bf 49}, 652 (1973).

\bibitem{Christenson:1964fg}
  J.~H.~Christenson, J.~W.~Cronin, V.~L.~Fitch and R.~Turlay,
  Phys.\ Rev.\ Lett.\  {\bf 13}, 138 (1964).

\bibitem{osci} Y. Fukuda et al. [SuperKamiokande Collaboration], Phys. Rev. Lett. 81 (1998) 1562; Phys. Rev. Lett. 82 (1999) 2644. Y. Fukuda et al., Phys. Lett. B 433, 9 (1998); Phys. Lett. B 436, 33 (1998).

\bibitem{Abdallah:2003xd}
  J.~Abdallah {\it et al.}  [DELPHI Collaboration],
  Eur.\ Phys.\ J.\ C {\bf 35}, 159 (2004)
  [hep-ex/0406010].


\bibitem{asm} M. A. Samuel, G. w. Li and R. Mendel, Phys. Rev. Lett. 67 (1991) 668;
M. A. Samuel, G. w. Li and R. Mendel, Phys. Rev. Lett. 69 (1992) 995.
Erratum; S. Eidelman and M. Passera, Mod. Phys. Lett. A 22 (2007) 159.

\bibitem{dsm}F. Hoogeveen, Nucl. Phys. B 341 (1990) 322; M. E. Pospelov and I. B. Khriplovich, Sov. J. Nucl. Phys. 53 (1991) 638 [Yad. Fiz. 53 (1991) 1030].

\bibitem{Acciarri:1998zc}
  M.~Acciarri {\it et al.}  [L3 Collaboration],
  Phys.\ Lett.\  B {\bf 426}, 207 (1998).



\bibitem{Acton:1992ff}
  P.~D.~Acton {\it et al.}  [OPAL Collaboration],
  Phys.\ Lett.\ B {\bf 281}, 405 (1992).

\bibitem{Heister:2002ik}
  A.~Heister {\it et al.}  [ALEPH Collaboration],
  Eur.\ Phys.\ J.\ C {\bf 30}, 291 (2003)
  [hep-ex/0209066].

\bibitem{Bernabeu:1994wh}
  J.~Bernabeu, G.~A.~Gonzalez-Sprinberg, M.~Tung and J.~Vidal,
  Nucl.\ Phys.\ B {\bf 436}, 474 (1995)
  [hep-ph/9411289].

\bibitem{Booth:1993af}
  M.~J.~Booth,
  hep-ph/9301293.

\bibitem{Georgi:2007ek}
  H.~Georgi,
  Phys.\ Rev.\ Lett.\  {\bf 98}, 221601 (2007)
  [hep-ph/0703260].

\bibitem{Moyotl:2011yv}
  A.~Moyotl, A.~Rosado and G.~Tavares-Velasco,
  Phys.\ Rev.\ D {\bf 84}, 073010 (2011)
  [arXiv:1109.4890 [hep-ph]].

\bibitem{Hektor:2008xu}
  A.~Hektor, Y.~Kajiyama and K.~Kannike,
  Phys.\ Rev.\ D {\bf 78}, 053008 (2008)
  [arXiv:0802.4015 [hep-ph]].


\bibitem{Cheung:2007ap}
  K.~Cheung, W.~-Y.~Keung and T.~-C.~Yuan,
  Phys.\ Rev.\ D {\bf 76}, 055003 (2007)
  [arXiv:0706.3155 [hep-ph]].

\bibitem{Ask:2008fh}
  S.~Ask,
  Eur.\ Phys.\ J.\ C {\bf 60}, 509 (2009)
  [arXiv:0809.4750 [hep-ph]].


\bibitem{Chatrchyan:2011nd}
  S.~Chatrchyan {\it et al.}  [CMS Collaboration],
  Phys.\ Rev.\ Lett.\  {\bf 107}, 201804 (2011)
  [arXiv:1106.4775 [hep-ex]].

\bibitem{Perl:1991gd}
  M.~L.~Perl,
  Rept.\ Prog.\ Phys.\  {\bf 55}, 653 (1992).

\bibitem{Gentile:1995ue}
  S.~Gentile and M.~Pohl,
  Phys.\ Rept.\  {\bf 274}, 287 (1996).

\bibitem{Pich:1997ym}
  A.~Pich,
  Adv.\ Ser.\ Direct.\ High Energy Phys.\  {\bf 15}, 453 (1998)
  [hep-ph/9704453].

\bibitem{Aad:2011kt}
  G.~Aad {\it et al.}  [ATLAS Collaboration],
  Phys.\ Rev.\ D {\bf 84}, 112006 (2011)
  [arXiv:1108.2016 [hep-ex]].

\bibitem{Chatrchyan:2011nv}
  S.~Chatrchyan {\it et al.}  [CMS Collaboration],
  JHEP {\bf 1108}, 117 (2011)
  [arXiv:1104.1617 [hep-ex]].

\bibitem{Aubert:2001tu}
  B.~Aubert {\it et al.}  [BABAR Collaboration],
  Nucl.\ Instrum.\ Meth.\ A {\bf 479}, 1 (2002)
  [hep-ex/0105044].

\bibitem{:2000cg}
  A.~Abashian, K.~Gotow, N.~Morgan, L.~Piilonen, S.~Schrenk, K.~Abe, I.~Adachi and J.~P.~Alexander {\it et al.},
  Nucl.\ Instrum.\ Meth.\ A {\bf 479}, 117 (2002).

\bibitem{Bernabeu:2007rr}
  J.~Bernabeu, G.~A.~Gonzalez-Sprinberg, J.~Papavassiliou and J.~Vidal,
  Nucl.\ Phys.\ B {\bf 790}, 160 (2008)
  [arXiv:0707.2496 [hep-ph]].

\bibitem{Inami:2002ur}
  K.~Inami [Belle Collaboration],
  eConf C {\bf 0209101}, TU10 (2002)
  [Nucl.\ Phys.\ Proc.\ Suppl.\  {\bf 123}, 74 (2003)]
  [hep-ex/0210035].

\bibitem{Banks:1981nn}
  T.~Banks and A.~Zaks,
  Nucl.\ Phys.\ B {\bf 196}, 189 (1982).

\bibitem{Grinstein:2008qk}
  B.~Grinstein, K.~A.~Intriligator and I.~Z.~Rothstein,
  Phys.\ Lett.\ B {\bf 662}, 367 (2008)
  [arXiv:0801.1140 [hep-ph]].


\bibitem{Nakayama:2007qu}
  Y.~Nakayama,
  Phys.\ Rev.\ D {\bf 76}, 105009 (2007)
  [arXiv:0707.2451 [hep-ph]].


\bibitem{Biggio:2008in}
  C.~Biggio,
  Phys.\ Lett.\ B {\bf 668}, 378 (2008)
  [arXiv:0806.2558 [hep-ph]].

\bibitem{Ibrahim:2008gg}
  T.~Ibrahim and P.~Nath,
  Phys.\ Rev.\ D {\bf 78}, 075013 (2008)
  [arXiv:0806.3880 [hep-ph]].

\bibitem{adtau-eff}  H. Fritzsch and Z.Z. Xing, Prog. in Part. and Nul. Phys. 45 (2000) 1; T. Huang, Z. H. Lin, X. Zhang, Phys. Lett. B 450(1999) 257; H. Fritzsch and Z. Z. Xing, Phys. Lett. B372 (1996) 265.


\bibitem{Ibrahim:2010va}
  T.~Ibrahim and P.~Nath,
  Phys.\ Rev.\ D {\bf 81}, 033007 (2010)
  [arXiv:1001.0231 [hep-ph]].

\bibitem{dm-hb}T. M. Aliev, K. Azizi, A. Ozpineci, Phys. Rev. D 77, 114006 (2008)[arXiv:0803.4420v2 [hep-ph]]; Amand Faessler, Th. Gutsche, M. A. Ivanov, J. G. Korner, V. E. Lyubovitskij, D. Nicmorus, K. Pumsa-ard, Phys. Rev. D 73, 094013 (2006) [arXiv:0602193v3[hep-ph]].

\bibitem{Baur:2004uw}
  U.~Baur, A.~Juste, L.~H.~Orr and D.~Rainwater,
  Phys.\ Rev.\ D {\bf 71}, 054013 (2005)
  [hep-ph/0412021].


\bibitem{Ibrahim:2010hv}
  T.~Ibrahim and P.~Nath,
  Phys.\ Rev.\ D {\bf 82}, 055001 (2010)
  [arXiv:1007.0432 [hep-ph]].

\bibitem{Hollik:1998vz}
  W.~Hollik, J.~I.~Illana, S.~Rigolin, C.~Schappacher and D.~Stockinger,
  Nucl.\ Phys.\ B {\bf 551}, 3 (1999)
  [Erratum-ibid.\ B {\bf 557}, 407 (1999)]
  [hep-ph/9812298].

\bibitem{Ibrahim:2011im}
  T.~Ibrahim and P.~Nath,
  Phys.\ Rev.\ D {\bf 84}, 015003 (2011)
  [arXiv:1104.3851 [hep-ph]].

\bibitem{Martinez:2008hm}
  R.~Martinez, M.~A.~Perez and O.~A.~Sampayo,
  Int.\ J.\ Mod.\ Phys.\ A {\bf 25}, 1061 (2010)
  [arXiv:0805.0371 [hep-ph]].

\bibitem{Soni:1992tn}
  A.~Soni and R.~M.~Xu,
  Phys.\ Rev.\ Lett.\  {\bf 69}, 33 (1992).


\end{thebibliography}
\end{document}